\documentclass[conference]{IEEEtran}
\IEEEoverridecommandlockouts
\usepackage{cite}
\usepackage{amsmath,amssymb,amsfonts}
\usepackage{algorithmic}
\usepackage{graphicx}
\usepackage{textcomp}
\usepackage{xcolor}
\usepackage{dblfloatfix}
\usepackage{array}
\pdfoutput=1
\newcolumntype{P}[1]{>{\centering\arraybackslash}p{#1}} 

\def\BibTeX{{\rm B\kern-.05em{\sc i\kern-.025em b}\kern-.08em
    T\kern-.1667em\lower.7ex\hbox{E}\kern-.125emX}}

\begin{document}

\title{Design and Simulation of a Hybrid Architecture \\for Edge Computing in 5G and Beyond}

\author{
    \IEEEauthorblockN{Hamed RAHIMI\IEEEauthorrefmark{1}\IEEEauthorrefmark{2}\IEEEauthorrefmark{4}\thanks{Hamed.Rahimi@etu.univ-st-etienne.fr}, Yvan PICAUD\IEEEauthorrefmark{1}, Salvatore COSTANZO\IEEEauthorrefmark{1}, \\Giyyarpuram MADHUSUDAN\IEEEauthorrefmark{1}, Olivier BOISSIER\IEEEauthorrefmark{4}, Kamal Deep SINGH\IEEEauthorrefmark{3}} \\
    \IEEEauthorblockA{\IEEEauthorrefmark{1}Orange Labs, Lannion, France}
\IEEEauthorblockA{\IEEEauthorrefmark{4} MINES Saint-Etienne, Saint-Etienne, France}
    \IEEEauthorblockA{\IEEEauthorrefmark{2}Université Jean Monnet, Saint-Etienne, France}
 \IEEEauthorblockA{\IEEEauthorrefmark{3} Télécom Saint-Etienne, Saint-Etienne, France}

}

\maketitle

\begin{abstract}
Edge Computing in 5G and Beyond is a promising solution for ultra-low latency applications (e.g. Autonomous Vehicle, Augmented Reality, and Remote Surgery), which have an extraordinarily low tolerance for the delay and require fast data processing for a very high volume of data. The requirements of delay-sensitive applications (e.g. Low latency, proximity, and Location/Context-awareness) cannot be satisfied by Cloud Computing due to the high latency between User Equipment and Cloud. Nevertheless, Edge Computing in 5G and beyond can promise an ultra-high-speed caused by placing computation capabilities closer to endpoint devices, whereas 5G encourages the speed rate that is 200 times faster than 4G LTE-Advanced. This paper deeply investigates Edge Computing in 5G and characterizes it based on the requirements of ultra-low latency applications. As a contribution, we propose a hybrid architecture that takes advantage of novel and sustainable technologies (e.g. D2D communication, Massive MIMO, SDN, and NFV) and has major features such as scalability, reliability and ultra-low latency support. The proposed architecture is evaluated based on an agent-based simulation that demonstrates it can satisfy requirements and has the ability to respond to high volume demands with low latency.
\end {abstract}


\section{Introduction}
Nowadays, Cloud Computing \cite{b1} is the most common option for many software vendors who are offering their application products and services over the Internet. Cloud Computing is a computing paradigm that offers on-demand services to the end-users through the pool of resources such as computation, data storage, analytics, and intelligence \cite{b2}.  The market of Cloud Computing has seen massive growth over the past decade. According to Statista\cite{b3}, Information technology (IT) services and business services will bring us more than 315 billion U.S. dollars revenue only in Europe, the Middle East, and Africa  in 2023. With the massive growth of mobile and IoT applications, the conventional centralized cloud computing comes upon several challenges, such as high latency, low Spectral Efficiency, and non-adaptive machine type communication due to the heavy load of data on the communication channels \cite{b4}. Moreover, the next-generation delay-sensitive applications, which fundamentally require quick response time and ultra-low latency, such as Autonomous Vehicle \cite{b5} and Augmented Reality (AR), add substantial stress on the network backhaul links providing connection to the cloud due to the enormous amount of data transferred\cite{b6}. To address such challenges, researchers suggest that we can process and analyze data just one level away from end-users. This concept constitutes a new computing paradigm called Edge Computing \cite{b7} that impressively reduces latency and bandwidth demands on network links by providing resources, services closer to the end-user in the Edge network. Edge Computing is developed to address the issue of high latency in delay-sensitive services and applications that are not properly handled within the Cloud computing paradigm \cite{b8}. In fact, the main difference between Edge computing and Cloud computing lies in the location of the servers. In other words, Edge Computing complements Cloud Computing by enhancing the end-user service to provide resources for delay-sensitive applications. Edge Computing has advantages such as fast processing, ultra-low latency, quick response time, real-time access to the network information, high bandwidth, mobility support, proximity to the user, and location awareness systems. On the other hand, Edge Computing also faces various inevitable challenges \cite{b9}. For instance, Cloud Computing is more scalable than Edge Computing, which in turn possesses limited resource and computing power as compared to the public cloud infrastructures or an enterprise data center\cite{b10}. However, Unlike Cloud computing, Edge computing takes advantage of distributed models for the distribution of servers as compared to Cloud computing that uses a centralized model\cite{b11}. Moreover, Cloud computing faces higher jitters due to the significantly high distance between User Equipment (UE) and the Cloud server that induces high latency in Cloud computing as compared to Edge computing\cite{b12}. Besides, due to the long distance between the end-user and the server, the probability of data end-route attacks is higher with Cloud computing than Edge computing\cite{b13}.\\

The fifth-generation (5G) networks\cite{b14} have been making a revolution into the world of science and technology. The 5G with speeds of up to 100 gigabits per second promises more mobile data speeds that are set to be 200 times faster than 4G LTE-Advanced. The 5G is the main guide for the growth of IoT applications by connecting billions of intelligent things to make the actual massive IoT\cite{b15} a reality. Based on the report of the International Data Corporation (IDC)\cite{b16}, 70\% of companies will be supported by the services of the global 5G by spending \$1.2 billion on the connectivity management solutions. Although 5G encompasses numerous technologies addressing low latency and reliable communications, it has three new main features to compare with LTE networks\cite{b17}. First, the amount of generated data is massively incomparable with the previous generation of cellular networks. Second, it makes reliable communications with ultra-low latency that satisfies the QoS requirement of delay-sensitive applications. Third, it supports heterogeneous environments to allow the interoperability of a diverse range of devices with different communication technologies. Besides, 5G networks support programmable and reconfigurable deployments of core network functions through Network Function Virtualization (NFV)\cite{b18} and Software-Defined Networks (SDN)\cite{b19}. Edge computing assists 5G networks for having a successful deployment in order to provide a large amount of broadband, low transmission latency, and a huge amount of communication connections simultaneously\cite{b20}.

To develop a solution for ultra-low latency applications, the authors believe that it is required to have an adapted architectural model for Edge Computing in 5G. This architecture should take advantage of suitable technologies to address the challenges of delay-sensitive applications. Besides, the proposed architecture model needs to demonstrate its capabilities before implementation. Moreover, to fulfill the proposed solution, it is required to consider a proper way for implementation. This paper aims to develop a solution for Edge Computing in 5G and beyond considering the requirements of delay-sensitive application. The proposed hybrid architecture has major features such as scalability and reliability. 

The rest of the paper is organized as follows. In Section II, we first review Edge computing and its deployment models as the state of the art. In Section III, Edge Computing in 5G and beyond is profoundly investigated. Besides, we characterize the requirements of the ultra-low latency use cases of Edge Computing in 5G. In section IV, we employ the characteristics of the requirements discussed in Section III and propose a hybrid architecture for edge computing in 5G and beyond. In Section V, we perform an agent-based simulation on three models to demonstrate that the proposed architecture has less latency compared to the other deployment models. In Section VI, We discuss the comparison between the proposed architecture and similar hybrid design architectures for Edge Computing. Finally, we review open issues of the proposed hybrid architecture and conclude the paper in Section VII.

\section{Literature Review}

Edge computing is a distributed data processing system consisting of numerous heterogeneous devices that communicate with the network in order to perform computing tasks\cite{b21}. Despite the Cloud Computing, Edge Computing places computation capabilities and the storage closer to endpoint devices to improve the capability of services and to reduce the latency of delay-sensitive applications\cite{b22}. Edge Computing is the most solid solution for many companies to decentralize or offload workloads to closer devices in order to bring various features to their services\cite{b23}. In the following, before reviewing the deployment models of edge computing, we investigate its benefits, challenges, and requirements.
\subsection{Benefits, Challenges, and Requirements of Edge Computing}
The most important benefits of Edge Computing are as follows\cite{b17}:
\begin{itemize}
  \item \textbf{Ultra-Low Latency-} The notable benefit of edge computing is the capacity to improve the performance of the system by reducing latency due to localization of data analyzing procedure, instead of transferring the collected information to far destinations under the traditional cloud computing architecture.
  \item \textbf{Scalability-} Although Edge Computing possesses limited resource and computing power as compared to the public cloud infrastructures, the building dedicated Edge Cloud is still rising due to the possibility of expanding computing capabilities flexibly of organizations at a low cost.
  \item \textbf{Reliability-} The operation of a system depends on the managing of processing functions.  In the case a system is provided with Edge Computing, the possibilities of the network impaction are negligible due to the closeness of Edge Cloud to the end-users.
  \item \textbf{Success Rate-} The probability of service shutdown is so rare in a system with a number of edge Clouds because Edge Computing offers unparalleled reliability by routing the data using multiple paths.
  \item \textbf{Security and Privacy-} The distribution of storage and processing carried across a wide range of Edge Clouds and End-Users provides a high level of security from vulnerable attacks. Besides, privacy issues are considered by Edge Computing due to storing and processing the generated data from IoT devices within nodes in the edge network.
\end{itemize}
Besides technical advantages, there are numerous financial efficiency and cost reduction with Edge Computing. For instance, there is no need to send large files to a cloud server where you would have to pay for storing them. According to Grand View Research\cite{b24}, the global value of the edge computing market in 2019 is \$43.4 billion. It has been estimated that the global size of the edge computing market will explode with an annual growth rate of 37.4 percent by 2027. Moreover, McKinsey\&Company has shown edge computing has a potential value of \$175 billion to \$215 billion in hardware by 2025\cite{b25}. This shows the importance of Edge Computing, especially for businesses to have better visibility into the operations of their business and better serve their customers.

Edge computing is renovating the way data is being handled through systems. Edge Computing, in order to promise faster processing right at the source of the data, can take advantage of several technologies such as Network Function Virtualization (NFV) plus Software-Defined Networking (SDN). In-depth, it is required to recognize the characteristics of Edge Computing alongside with its enabling requirements in order to study its deployment models. Edge computing has five main characteristics as follows\cite{b26}: 
  \begin{itemize}
  \item \textbf{Low latency and proximity:} The availability of the computational resources in the local proximity reduces the response delay time experienced by devices as compared to reaching the traditional cloud. Also, it allows users to leverage the network information in order to make offloading decisions. Moreover, the service provider can leverage the information of users by extracting and analyzing device information and the user's behavior to improve their services.
	\item \textbf{The Dense Geographical Distribution:} it improves the accuracy of the system by facilitating location-based mobility services with real-time analytics to the network administrators without traversing the entire WAN. 
  \item \textbf{Location awareness:} As the number of devices is rapidly increasing, Edge computing supports mobility, such as the Locator ID Separation Protocol (LISP), to communicate directly with devices. Edge computing is location-aware and highly supports mobility by enabling edge servers to collect and analyze the generated data based on the geographical location of UEs.
  \item \textbf{Context awareness:} Context-awareness is defined interdependently to location awareness and is a characteristic of mobile devices that enables edge servers to collect network context information. Edge computing takes offloading decisions and accesses the Edge services based on the context information of the mobile device.
  \item \textbf{Heterogeneity:} The network heterogeneity refers to the diversity of communication technologies that affect the performance of Edge services. The successful deployment of Edge computing handles interoperability issues and covers varied platforms, architectures, infrastructures, computing, and communication technologies used by the Edge computing elements such as End-Devices, Edge-Servers, and Networks.
\end{itemize}

Considering the characteristics of Edge Computing, there are five major enabling requirements for Edge computing that are classified as follows. 1) Real-time Interaction: Realtime interaction enables the use of edge computing over cloud computing to improve QoS and to support ultra-low latency applications such as 3D Gaming that demands a very high speed of data communication. 2) Local processing: it creates an opportunity for more capacity of processing data and requests by high speed communication between edge servers. 3) High data rate: it transmits the massive amount of data generated to edge clouds. 4) Dynamic Billing Mechanism: it refers to the ability of a service provider to manage its variables when an end-user is facilitated through their roaming Edge services. This mechanism depends on various network parameters such as bandwidth and latency to respond to several end-users who may request specific resources from the cloud through the Edge Cloud. 5) high availability: it ensures the availability of the cloud services at the edge.
\subsection{Deployment models of Edge Computing}
 In this section, we investigate three important deployment models of Edge Computing: Fog Computing, Cloudlet, and Multi-access Edge Computing (MEC)\cite{b6}\cite{b7}\cite{b8}. 
\subsubsection{Fog Computing}
Fog computing\cite{b27} has been introduced by Cisco in 2012. This model enables applications to run directly at the edge of the network through billions of smart connected devices. Fog computing is a cooperation-based edge cloud, wherein heterogeneous end-users share their resources in order to deliver services and applications. The OpenFog Consortium\cite{b28}, which has been founded by ARM, Cisco, Dell, Intel, Microsoft, and Princeton University in November 2015, defines fog computing as a system-level horizontal architecture that distributes resources and services of computing, storage, control, and networking anywhere from the cloud to end-users. In the architecture of Fog Computing various tools have been provided to distribute and manage resources and services along with end users and across networks. However, servers, applications, and small clouds are placed at the edge in other architectures. The difference between Edge Cloud and Fog is that Fog interacts with industrial gateways and embedded computer systems on a local area network, with various operations depending on the cloud, whereas edge performs on embedded computing platforms directly interfacing to sensors and controllers. 

\subsubsection{Cloudlets}
 The cloudlets \cite{b29}  or “a data center in a box” is a small-scale cloud Data Center (DC) that is located at the edge of the internet and offers a bunch of resources and services for computing, storage, control, and wireless LAN connectivity towards the edge. Cloudlets solve the latency problem and the end-to-end response time between end-users and associated Cloud by using the computer resources available in the local network. A cloudlets is a cluster of secure computers with high performance and capability of computing that are connected to the internet and ready to use by nearby end-users. The main idea of the cloudlets lies behind supporting resource-intensive and interactive applications by providing powerful computing resources to end-users with lower latency. Cloudlets represent the middle layer of the 3-level hierarchy architecture\cite{b29}, mobile device layer, cloudlets layer, and cloud layer, to attain lower response time. In the Cloudlets architecture, User Equipments (UEs) have access to resources of cloudlets through a one-hop high-speed wireless LAN.

\subsubsection{Multi-access Edge Computing (MEC)}
 The concept of Mobile Edge Computing\cite{b30}, which has been supported by Huawei, IBM, Intel, Nokia Networks, NTT DoCoMo, Vodafone, and other more companies, has been introduced by European Telecommunications Standards Institute (ETSI)\cite{b31} in September 2014. In this Model, mobile users can utilize computing services from the base station. In the MEC World Congress 2016, MEC ISG has renamed mobile edge computing as multi-access edge computing in order to reflect the growing interests of non-cellular operators. MEC applications are running as VM on top of the virtualization infrastructure and can interact with the mobile edge platform to perform certain support procedures related to the life-cycle of the application.

To compare these three models, we should indicate that the Cloudlet and the Multi-access Edge Computing are designed to provide the services to end-users with the local resources. However, Fog especially relies on the hardware that possesses computational capabilities along with the normal functionality of the device such as router and switches. Telecommunication operators typically use MEC that is recognized as a key model that satisfies 5G requirements\cite{b32}, while Cloudlets and Fog are independent of operators and are typically hosted by B2B clients. Each implementation model has its characteristics that are proper for precise use cases. As shown in table I, we compare the deployment models of Edge Computing with each other considering the characteristics (e.g. Context-Awareness) and requirements (e.g Real-time Interaction and Security) of the deployment models. We can conclude from the table that MEC and Cloudlet are more suitable for use cases that require more resources for computation. However, Fog Computing is better for cases that need fewer resources and lower latency.

\begin{table*}[h!]
\begin{center}
\caption{DEPLOYMENT MODELS OF EDGE COMPUTING}
\begin{tabular}{ |P{2.75cm}|P {5.75cm}| P{3.75cm}| P{4.75cm} | } 
\hline
Type of Implementation & Fog Computing & MEC & Cloudlets \\ 
\hline
Organizations & ARM, Cisco, Dell, Intel, Microsoft, and Princeton University founded OpenFog Consortium & AT\&T, Huawei, Intel, Vodafone, etc. supports ETSI MEC & Vodafone, Intel, Huawei, and Carnegie Mellon University launched OEC\\ 
\hline
Real-time Interaction & High & Medium & Medium \\ 
\hline
Computation Power & Medium & High & High \\ 
\hline
Power Consumption & Low & High & Medium \\ 
\hline
Coverage & Low & High & Low \\ 
\hline
Server Density & Medium & Low & High \\ 
\hline
Context-awareness & Medium & High & Low \\ 
\hline
\end{tabular}
\end{center}
 \end{table*}
 
\section{Edge Computing in 5G and Beyond}
 5G has been appreciated as the most important wireless cellular network that is expected to satisfy the requirements of next-generation networks\cite{b33}. There are three main characteristics in 5G networks that are not observed in the previous generations\cite{b34}\cite{b17}. As the first characteristic, 5G networks generate a massive amount of data. Secondly, although 4G impose QoS requirements, 5G networks provide more QoS granularity to support delay-sensitive, interactive and ultra-low latency applications. Third, 5G networks support heterogeneous environments by empowering the interoperability of a diverse range of networks and communication protocols. All these original characteristics of 5G are the result of three main brand-new technologies that provide higher capacities to networks\cite{b17}. The First is mmWave communication, which uses high-frequency bands like 30 GHz to 300 GHz to provide high bandwidth at a minimum of 11 Gbps. Second, the deployment of small-cells that enables devices to communicate through mmWave to decrease interference and transmission range. Finally, the massive multiple-input multiple-output (MIMO) that empowers base stations (BSs) in order to use a considerable number of antennas to enable beamforming and reduce interference by supporting parallel communication. 

\subsection{Components of 5G and Beyond}
5G networks consist of four components: 5G Radio Access Network (RAN), 5G Core Network (CN), 5G Transport Network, and 5G Interconnect Network. 5G has split the Evolved Packet Gateway (EPG) of 4G, the Ericsson's implementation of the 3GPP S-/P-GW, into User Plane Function (UPF) and Session Management Function (SMF) as a data plane function and a control plane function, respectively. The UPF handles the critical data plane and the packet processing between RAN and the data network such as the Internet\cite{b35}. 5G Radio Access Technology (RAT) is the underlying physical connection method for the 5G communication network. 5G Multiple Radio Access Technology (Multi-RAT) RAN\cite{b36} is an access network that consists of various cellular networks and communication protocols such as Bluetooth, Wi-Fi, etc. 5G gNB is a node of 5G RAN connected to the 5G CN. 5G gNB\cite{b37} is the next generation NodeB that is a radio base station similar to what was called NodeB in 3G-UMTS and eNodeB or eNB in 4G-LTE. 5G gNB architecture is split into Central Unit (CU) and Distributed Unit (DU), which are the control plane and data plane of the 5G RAN, respectively. 5G CU is a logical node that controls the gNB functions such as Session Management, Transfering the user data, Mobility control, and sharing RAN.  5G CU can be hosted in a regional cloud data center or collocated with a 5G DU.  5G DU is a logical node that operates based on the option of a lower-layer split (LLS). The operation of 5G DU is controlled by 5G CU and it is hosted in an Edge cloud data center connected to the radio via a packet interface known as enhanced Common Public Radio Interface (eCPRI). Besides, 5G CU could be split into 5G CU Control Plane (5G CU-C) part and 5G CU User Plane (5G CU-U), and be implemented in different locations\cite{b38}. Researchers around the world have started to investigate 6G networks as beyond 5G networks that are being commercialized. 6G is expected to have a profound impact on the intelligence process of communication developments to deliver high Quality of Service and energy efficiency by adjusting new architectural changes consisting of intelligent connectivity, deep connectivity, holographic connectivity, and ubiquitous connectivity\cite{b39}. 5G and beyond empowers Edge Computing aiming to move the computation away from data centers towards the edge of the network. 
\subsection{Objectives of Edge Computing in 5G and Beyond}
There are five main objectives that edge computing in 5G and beyond pursues as follows\cite{b40}.
  \begin{itemize}
  \item \textbf{Enhancing data management:} it improves the capability of a system to handle a massive amount of data that is generated by devices.
  \item \textbf{Enhancing the Quality-of-Service (QoS):} it improves the QoS in order to support various requirements of QoS and to improve the Quality-of-Experience (QoE). Service providers are required to take a holistic view of their subscribers with contextual information such as preferences and interests. Afterward, the personalized information could be used in order to enhance their QoE or for attracting new customers.
  \item \textbf{Network demand Estimation:} it predicts the required network resources in order to support the demands of a network in local proximity and to provide an optimal allocation of resources to handle the local demands of the network.
  \item \textbf{Dynamic resource allocation:} in order to allocate resources of the network such as bandwidth in an efficient way.
  \item \textbf{Location awareness Management:} edge servers are distributed geographically in order to support location-based services by inducing and tracking the location of devices. 
  \item \textbf{Resource Management:} it enhances the optimization of network resource utilization to improve network performance.
\end{itemize}

\subsection{Technologies of Edge Computing in 5G and Beyond}
5G and beyond assist the next generation of IoT systems by satisfying the four key requirements of edge computing discussed in the previous section with major technologies that enable edge computing in 5G and beyond. These technologies are as follows.
\begin{itemize}

 \item \textbf{Cognitive Radio and Intelligent Reflecting Surfaces (IRSs):} Cognitive Radio\cite{b41} allows us to reach the high spectrum demand and efficient spectrum scarcity of ultra-low latency applications and services of massive IoT connections.
Besides, Intelligent Reflecting Surfaces (IRSs)\cite{b42} increase spectrum and energy efficiencies by tuning to the wireless environments. These two are promising technologies for 6G wireless communication.

 \item \textbf{Network Function Virtualization (NFV):} NFV involves the implementation of network functions as software modules that can run on general-purpose hardware. NFV enables edge devices to provide computing services and operate network functions by creating multiple Virtual Machines (VMs). By this approach, it is not necessary to dedicate specific hardware to run for each network functions and services.
  \item \textbf{Software-Defined Networks (SDN):} SDN is a network architecture that separates a network into control and data planes to provide flexible and agile networks, which helps to simplify network management and deploy new services. In fact, SDN complements NFV by decoupling the management or control plane from the data plane over which data packets are forwarded. SDN enables easier and more flexible management of networks through abstractions and a logically centralized controller that handles policy and forwarding decisions.
  \item \textbf{Massive Multiple-Input Multiple-Output (MIMO):} Massive MIMO \cite{b43} increases the number of antennas at the transmitter and the receiver by deploying multiple antenna elements. This is following the Shannon theorem in which we can increase the signal-to-noise ratio without increasing the transmission power, which leads us to enhanced network capacity and energy efficiency. Using massive MIMO, we can offload tasks between edge servers simultaneously in order to reduce latency and energy consumption.
  \item \textbf{Computation Offloading:} Computation offloading \cite{b23}\cite{b44} is a method that transfers resource-intensive computations from a device or edge server to the resource-rich nearby device or edge server. The most important part of computation offloading is to decide whether to offload or not, whether full or partial offloading is applicable, and what and how the computation could be offloaded. The offloading decision depends on the application model. Based on various techniques and algorithms, the decision engine determines whether to offload or not. The majority of algorithms aim to minimize the energy consumption at the mobile device, subject to the execution delay acceptable by the offloaded application, or to find an optimal tradeoff between these two metrics.
  \item \textbf{5G NR (New Radio):} 5G NR\cite{b45} is a new Radio Access Technology (RAT) developed by 3GPP that provides a connection to a diverse range of devices to achieve low latency and scalable networks.
  \item \textbf{Device-to-Device (D2D) communication:} D2D communication\cite{b46} allows direct communication between neighboring devices using ad-hoc links without passing through base stations. D2D communication improves energy efficiency and spectrum utilization of the system. Moreover, devices can offload tasks and computations to edge servers using D2D communication in order to improve the computational capabilities of UEs. Besides, D2D guarantees better QoS for service provisioning by supporting a large number of information interactions and providing a higher transmission data rate.
  \item \textbf{Cross-Layer Communication:} To enable the communication between different networks, we use network gateways, which are computer systems or devices that act as a conversion and translate various communication protocols, data formats, and network architectures. The communication between different devices with different communication protocols and data structures in a network gateway is defined as Cross-Layer Communication. This solution does not impose any modifications on the hardware of devices, and the whole design is transparent to the upper-level wireless communication systems\cite{b47}.
\end{itemize}

\subsection{Use cases of Edge Computing in 5G and Beyond}
To develop a solution for Edge Computing in 5G, we first need to recognize the requirements of ultra-low latency applications. In this regard, it is necessary to study applications of 5G Edge Computing to identify and classify their requirements. Various use cases are relying on 5G edge computing due to its capabilities on real-time interaction, local processing, high data rate, and high availability\cite{b48}. The most important use cases of 5G Edge Computing are as follows.
 \begin{itemize}
  \item \textbf{Remote Surgery:} The remote diagnostics and the remote surgery, as well as monitoring of patient vital signs, are of the most significant use cases of Health Care applications in 5G Edge Computing. These kinds of applications require an ultra-low latency with the accuracy of nanoseconds to react in real time during a surgery. Besides, these use cases demand trustworthy privacy as they carry the vital health information of a person. Further, it requires secure communication between Cloud Edge for processing data because if a wrong data gets injected into the system during surgery, it could bring a total disaster.
  \item \textbf{3D Gaming:} 3D Gaming, alongside multimedia services such as streaming HDTV or 3D TV, is the most significant use case of Entertainment applications in 5G Edge Computing. These use cases demand trustworthy privacy and security, and wide bandwidth to get updated in real-time with the nanosecond latency.
  \item \textbf{Virtual Reality (VR) and Augmented Reality (AR):} AR and VR use cases, such as streaming video contents on the virtual reality glasses, are the most viral use cases of Edge Computing that require large-scale bandwidth, ultra-low latency, and trustworthy security and privacy. 
  \item \textbf{Smart Factory:} Future factories are equipped with smart machines to improve safety and productivity. These factories require a scalable structure that has a very low latency for processing data during the industrial processes. Besides, the privacy and security of these factories are of the most vital requirements in Industry 4.0.
  \item \textbf{Autonomous Vehicles:} Autonomous Vehicles alongside with Internet of Vehicles (IoV), whereby drivers can collect information from traffic and process them in a real-time manner to avoid accidents or design new services, are the most significant use cases of the next generation transportation applications. These use cases require a scalable, trustworthy, and secure structure that is aware of its location.
\end{itemize}
\subsection{Deployment Model for Edge Computing Use Cases in 5G}
In Table II, we investigate the role of the characteristics and requirements discussed in the previous section for use cases of Edge Computing in 5G and Beyond. There are seven factors\cite{b49} such as Bandwidth, Ultra-Low Latency, and etc. that have been explored based on their necessity from crucial to important and incidental. For Instance, Location awareness is crucial for Autonomous Vehicles while it is only important for Remote Surgery. 
\begin{table*}[t]
\begin{center}
\caption{The requirements of Edge Computing use cases in 5G and Beyond}
\begin{tabular}{ | P{4cm} | P{3cm}| P{2cm}| P{2cm} | P{2cm}| P{2cm} | } 
\hline
Requirements/Applications & Autonomous Vehicles & Smart Factory & AR/VR & 3D Gaming & Remote Surgery\\ 
\hline
Bandwidth & important & important & crucial & crucial & important\\ 
\hline
Ultra-Low Latency & crucial & crucial & crucial & crucial & crucial\\ 
\hline
Extensibility & important & incidental & important & important & incidental \\ 
\hline
Context/Location Awareness & crucial & important & important & important & incidental\\ 
\hline
Power Consumption & crucial & crucial & crucial & crucial & crucial \\ 
\hline
Scalability & important & important & important & important & important \\ 
\hline
Privacy and Security & crucial & crucial & crucial & important & crucial\\ 
\hline
\end{tabular}
\end{center}
 \end{table*}
Furthermore, we determine the relationship between the mentioned requirements and the implementation features of the Edge Computing deployment models. As shown in Table III, this relationship links the characteristics of Edge Computing Use Cases in 5G (e. g. Bandwidth, Scalability, etc.) and fundamental features of its deployments model (Coverage, Real-Time Interaction, etc.). For instance, Scalability is highly related to coverage while it is less related to Real-Time Interaction. Table III can guide us to develop the most suitable solutions for the mentioned ultra-low latency use cases of Edge Computing in 5G and Beyond.
\begin{table*}[t]
\begin{center}
\caption{The requirements of Edge Computing use cases in 5G and the features of the Edge Computing deployment models}
\begin{tabular}{ | P{5cm} | P{2cm}| P{3cm} | P{3cm}| P{2cm} | } 
\hline
Requirements/Deployment Features & Coverage & Computational Capability & Real-time Interaction & Proximity\\ 
\hline
Bandwidth & Highly Related & Mostly Related & Highly Related & Highly Related \\ 
\hline
Ultra-Low Latency & Highly Related & Mostly Related & Highly Related & Highly Related \\ 
\hline
Extensibility & Mostly Related & Highly Related & Mostly Related  &  Highly Related\\ 
\hline
Context/Location Awareness & Mostly Related & Mostly Related & Highly Related &  Less Related\\ 
\hline
Power Consumption & Highly Related & Highly Related & Mostly Related & Mostly Related \\ 
\hline
Scalability & Highly Related & Mostly Related & Less Related &  Mostly Related\\ 
\hline
Privacy and Security & Mostly Related & Less Related & Mostly Related & Highly Related\\ 
\hline
\end{tabular}
\end{center}
 \end{table*}
In this section, we characterized the use cases of 5G Edge Computing in Table II and determined the relationship between the mentioned characteristics and the implementation features of the Edge Computing deployment models in Table III. Moreover, in the previous section, we discussed the requirements and characteristics of deployment models in Edge Computing and compared its deployment models with each other considering those characteristics and requirements in Table I.  Now, there is a link between deployment models of Edge Computing and the ultra-low latency use cases of 5G Edge Computing using the tables and information above. In the next section, we will propose an architecture that fulfills the requirements of Edge Computing use cases in 5G and Beyond.
\section{The Proposed Hybrid Architecture}
In this section, we are going to find the relationship between the deployment models of Edge Computing and the use cases of Edge Computing in 5G and Beyond. Using this relationship, we will propose a scalable architecture that fulfills the requirements of Edge Computing in 5G and Beyond. As we discussed in the previous section and Table III, every use case has its own particular requirements and characteristics. Also in Table I, we discussed the characteristics of the deployment models of Edge Computing. Using Table I and Table III, we can link the deployment models of Edge Computing to the use cases of 5G Edge Computing. For Instance, according to Table II and Table III, the requirements of Smart Factories are real-time interaction, to support extensible coverage, and scalability. Using Table I, we can find out that the Fog Computing paradigm is able to respond properly to the requirements of real-time interaction. However, in order to support extensible coverage with scalability, MEC can perform much more productively. In such a situation, the authors propose to use the integration of Fog Computing with MEC that can take advantage of low-layer processing paradigms such as performing Cross-Layer Communication (CLC) in Gateways in order to fulfill all requirements of Smart Factories. We use logic above and find suitable deployment models for the use cases of 5G Edge Computing listed in Table IV.
\begin{table}[h]
\begin{center}
\caption{The suitable models for the use cases of 5G Edge Computing}
\begin{tabular}{ | P{3cm} | P{4cm} | } 
\hline
Use Case &	Suitable Implementation Model\\
\hline
Autonomous Vehicles &	Fog Computing + MEC + D2D\\ 
\hline
Smart Factory &	Fog Computing + MEC + CLC \\ 
\hline
AR/VR & 	MEC/Cloudlets+ Fog Computing \\ 
\hline
3D Gaming &	MEC + Cloudlets \\ 
\hline
Remote Surgery	& Fog Computing/Cloudlets + MEC \\
\hline
\end{tabular}
\end{center}
 \end{table}

\begin{figure}[b]
    \centering
    \includegraphics[scale=0.9]{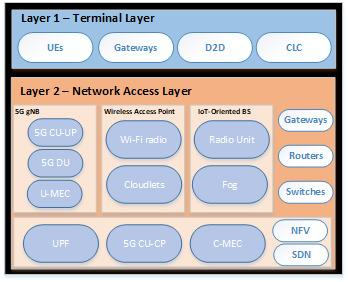}
    \caption{\textbf{The Proposed Hybrid Architecture for 5G Edge Computing}}
    \label{fig:4}
 \end{figure}

\begin{figure*}[h]
    \centering
    \includegraphics[scale=0.6]{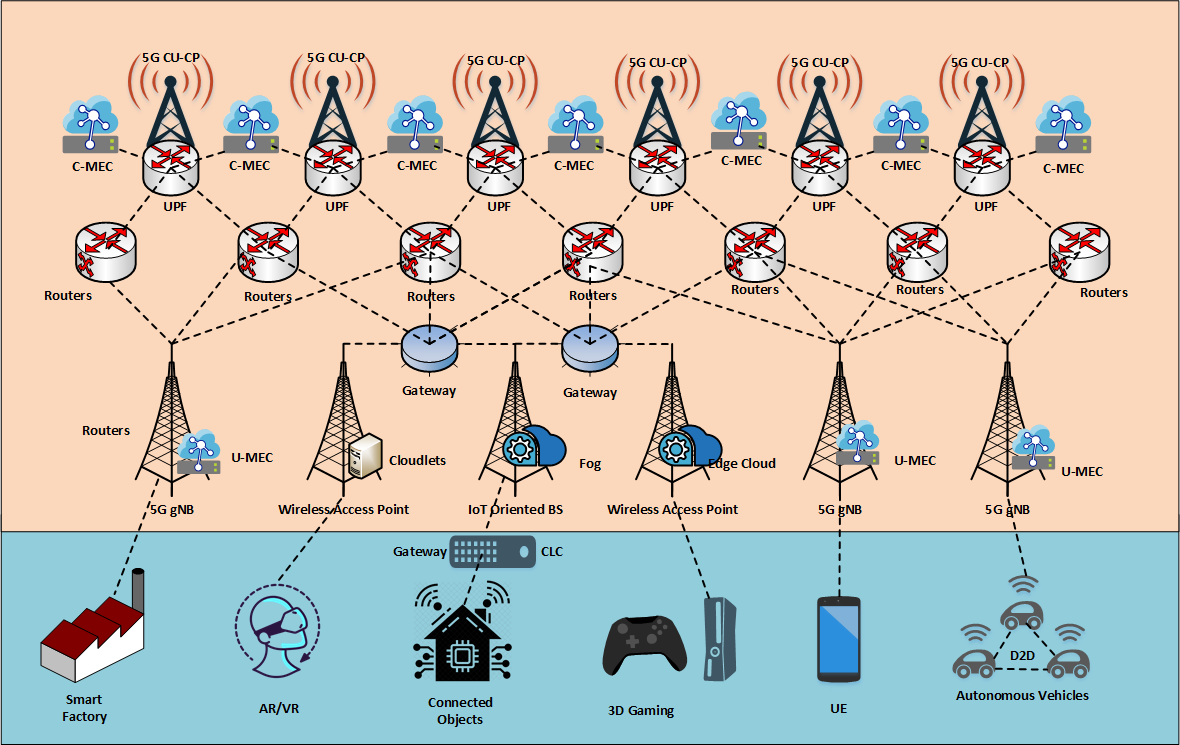}
    \caption{\textbf{The Deployment Model of The Proposed Architecture}}
    \label{fig:5}
 \end{figure*}

Using Table IV, we can conclude that the most suitable solution that covers uses cases of 5G Edge Computing that fulfill their requirements is the hybrid solution that supports the data pre-processing using CLC and D2D communication to enable collaborative computing that performs tasks in more than single computing platforms or paradigms. In this paper, the term "Hybrid Architecture" points out to an architecture that is integrated with MEC and Fog, and Cloudlets. As shown in Fig \ref{fig:4}, we are going to propose a hybrid architecture consisting of two layers before the core network that are Terminal Layer and Network Access Layer. As the deployment of the proposed architecture has been shown in Fig \ref{fig:5}, the proposed architecture is heterogeneous, scalable, and supports the requirements of 5G Edge Computing use cases.

\subsection{Terminal Layer}
The Terminal layer consists of UEs and gateways. In this layer, customers/UEs get connected to the 5G network directly by 5G transmitters or through a gateway. The role of this layer is to act as an Edge Platform as a Service (E-PaaS)\cite{b50} in order to achieve two objectives: The First is to take advantage of the capability of our customers’ hardware to preprocess the information and to perform uncomplicated analysis. The second is to use 5G in order to enable D2D communication between UEs. Not so long, vehicles will be autonomous and drive by themselves. Therefore, it is required to connect vehicles through a very low latency technology that can communicate in less than one microsecond. D2D Communication is a technology, which plays an important role in Connected-Vehicles. In this regard, we can consider D2D communication in the deployment of our platform. By exchanging and sharing the data between UEs through D2D communication,  we can have access to more data for processing and also creating innovative services. Furthermore, the E-PaaS should be able to use other common communication protocols\cite{b51}  such as Bluetooth for D2D communication between UEs in case one of them is not equipped with 5G. In this way, the E-PaaS can form an ad-hoc network that consists of various communication protocols. For instance, Google has designed Cloud IoT Edge, which is a software stack that runs on Android and Linux OS to extend AI and Data Analysis on the very low-level Edge\cite{b52}. Another important aspect of the E-PaaS platform is that it should be able to perform on gateways for utilizing Cross-layer Communication (CLC). Up to now, the common communication protocol of most IoT systems has been based on Wi-Fi, Bluetooth LP, ZigBee, and, recently, LPWANs such as Lora and Sigfox. To adapt 5G with this generation of IoT facilities, researchers have proposed Gateways to collect data of these systems, store and send them to a 5G gNBs, or directly to an individual UPF integrated with 5G CU-C. In such a situation, we propose to use Cross-Layer Communication (CLC) within these Gateways due to the enormous valuable information and services that could be retrieved from collecting and analyzing data through CLC in gateways. 
\subsection{Network Access Layer}
As discussed earlier, we propose to use a hybrid structure for edge computing that consists of two components toward a dynamic ultra-fast arrangement that is a combination of MEC, Fog, and Cloudlet. The first component is located on base stations and could be MEC, Fog, or Cloudlets considering the type of UE's communication protocol. In this regard, we have considered that the deployment of the proposed architecture is equipped with 5G Multi-RAT networks that combine several Radio Access Technologies to deliver the service to UEs. In this section, we classify the base stations into 3 categories: 1) 5G gNB 2) Wireless Access Point 3) IoT-Oriented Base Stations. For Wi-Fi-based BS such as Wireless Access Points, we propose to use Cloudlets as the deployment model of Edge Computing due to its characteristics that offer services for wireless LAN connectivity and cover the requirements of ultra-low applications such as VR/AR that are sensitive to delay and require to use real-time interactions. For IoT-Oriented protocols such as Lora and Sigfox, we suggest continuing using Fog Computing that is currently the most common deployment model for their use cases due to the high capability of processing information and the ability to run Cloud applications on the edge level. For 5G networks, as discussed in Section III, in the architecture of 5G gNB, 5G CU consists of 5G CU-CP and 5G CU-UP. In this Layer, the generated data from Terminal Layer will be provided to 5G gNB through the 5G Radio Unit (RU) and 5G gNB process the data by the MEC called U-MEC and is equipped with 5G CU-UP. The U-MEC is a type of MEC that has as high computation power as Fog Computing and also various tools to provide services along with UEs and across the network. In the proposition of U-MEC, we are mainly focused on the speed and capacity of data processing, whereas the second component is focused on distributing and managing the resources.

The second component is the core of the proposed 5G Edge Computing architecture that is located in a UPF integrated with 5G CU-CP and MEC Hosts called C-MEC. C-MEC is a type of MEC that is responsible to manage resources and data processing of the network. C-MEC is suitable for the core SDN and NFV due to its significant coverage support,  context-awareness support, and high server density. For the second component, we propose to use the MEC that is designed by the European Telecommunication Standard Institute. ETSI has standardized 5G architecture for plenty of telecommunication companies such as Vodafone and has considered various technologies NFV and SDN in its structure. ETSI has provided his architecture deployment in the Network Functionality Virtualization (NFV) environment, which allows MEC applications to run as a VM on top of the virtualization infrastructures. Besides, ETSI has designed an architecture for the SDN model that is responsible to control the network by Switches/Routers that are distributed in order to modify the network for computation offloading and resource allocation through Software Defined Networks (SDN) technology.

The proposed hybrid architecture satisfies the requirements of 5G Edge Computing use cases and fulfills the demands of 5G-IoT services. The hybrid architecture is developed based on the novel technologies, which empower the 5G Edge Computing paradigm that is more sustainable and scalable than existing architectures and has main features such as scalability, heterogeneousness, and ability to respond to the high volume of demands. There are numerous criteria to evaluate the proposed hybrid architecture and demonstrate its efficiency. In the next section, we will simulate the proposed architecture and will consider the latency as a significant criteria for the demonstration.

\section{Simulation and Analysis}
In order to demonstrate the efficiency of the proposed architecture, we have simulated the proposed architecture and considered the latency as vital criteria due to the requirements of delay-sensitive applications discussed in previous sections. Considering the time characteristics of information, there are three main categories for data\cite{b17}. The first type is hard real-time data that has a strict predefined latency. The second is soft real-time data that can tolerate predefined and bounded latency. The last is non-real-time data that is not time-sensitive and can tolerate latency. In this simulation, we have considered that all the data are hard real-time and cannot tolerate latency to meet the simulation with practical application.

In this regard, we have applied agent-based simulation\cite{b53}\cite{b54} to the proposed architecture and have modeled it as a Multi-Agent System (MAS)\cite{b55}. A multi-agent system is a distributed system composed of multiple interacting intelligent agents. Agent-based modeling is a powerful simulation technique, which has a collection of autonomous decision-making entities called agents that are organized in an artificial environment\cite{b56}. The main reason that we have used agent-based simulation for Edge Computing is that Edge Computing has a distributed architecture, and agent-simulation is a suitable solution for the modeling and execution of the distributed components of the architecture where every component individually makes decisions based on a set of rules and the local state of the system.

\begin{figure*}[h]
    \centering
    \includegraphics[scale=0.6]{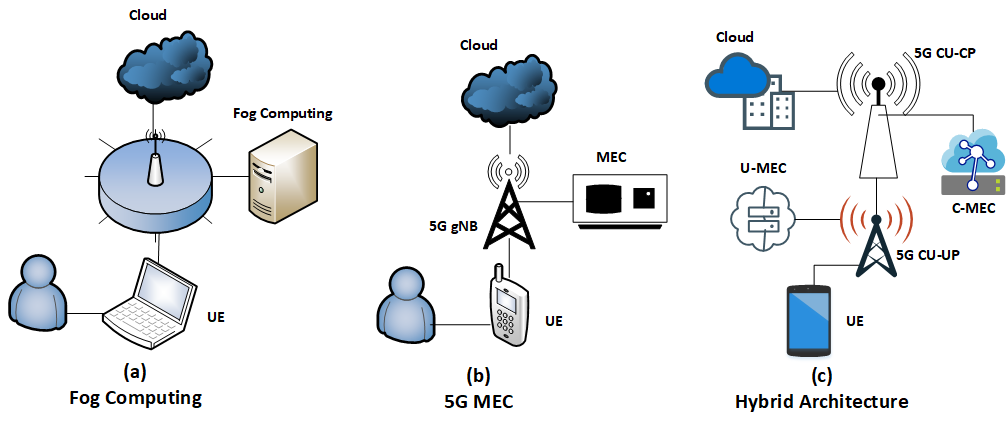}
    \caption{\textbf{The Paradigms of the Simulated Architectures}}
    \label{fig:6}
 \end{figure*}

As shown in Fig \ref{fig:6}, we have performed three agent-based simulations on three deployment models: 1) Fog Computing 2) 5G MEC 3) the proposed hybrid architecture. The strategy is to compare the latency of three simulations with each other. In this regard, we first have compared the latency of the Fog Computing and latency of the 5G MEC to show how 5G impacts on the latency and decreases it up to 50 times. Next, we have compared the latency of the 5G MEC and the latency of the proposed hybrid architecture to demonstrate the benefits of the proposed architecture. The simulation consists of nine types of agents: UE, Base Station (BS), 5G gNB, 5G CU-CP, 5G CU-UP, Fog, Edge Cloud, C-MEC, U-MEC, and Cloud. In this regard, we have used the osBrain\cite{b57} environment, which is a Python-based multi-agent system framework that allows us to have efficient and asynchronous communication with various communication patterns such as request-reply, push-pull, and publish-subscribe.

In the first model, we simulate a simple Fog Computing consisting of four types of agents: UE, BS, Fog, and Cloud as shown in Fig \ref{fig:6}(a). The UE continuously sends data to BS that is responsible to distribute the generated data between Fog or Cloud for computation. In this simulation, the latency has been measured based on the time between sending the raw data and receiving the processed information. As shown in (1), Latency is defined as \(T_{total}\), which is calculated through the sum of the Transmission Time  \(T_{T}\) and Processing Time \(T_{P}\) in Fog and Cloud. The Transmission Time is equal to the sum of the time that data is sent and received from UE to BS, a part of it from BS to Fog, and the rest of it from BS to Cloud in the case that the volume of data is high or the time we reserve the Fog for more sensitive application. The time that data is sent from UE to BS is defined \(T_{UE}=\frac{\alpha V_{Data}}{S_{UE}}\), in which \(V_{Data}\) is the amount of data, \(\alpha\) is the motion effect of UE considered as a randomized ratio between 0.7 and 1 in the speed rate, and \(S_{UE}\) is the speed of UE in sending the data to BS that supports 4G LTE-Advanced with the speed rate of 37 megabytes per second and is connected to the Fog and Cloud through the internet. The distance between UE-to-BS and BS-to-fog is set to a single hop, which impacts the latency considering the ideal situation for the communication part. The time that a part of data is sent from BS to Fog is defined \(T_{Fog}=\frac{V_{Fog}}{S_{BS}}\), in which \(V_{Fog}\) is the amount of data sent to Fog and \(S_{BS}\) is the speed of BS in sending the data to Fog. The time that the rest of data is sent from BS to Cloud is defined \(T_{Cloud}=\ln{H}*\frac{V_{Cloud}}{S_{BS}}\), in which \(V_{Cloud}\) is the amount of data sent to Cloud, \(S_{BS}\) is the speed of BS in sending the data to Cloud, and \(\ln{H}\) is the nonlinear effect of H number of hops between BS and Cloud, which is considered equal to 10 hops, due to the power loss and the capacity of the communication channels. On the other hand, Processing Time is equal to the time that data is processed by Fog and Cloud. The Processing Time is defined \(T_{P}=\frac{V_{Fog}}{C_{Fog}}+\frac{V_{Cloud}}{C_{Cloud}}\) in which \(V_{Fog},C_{Fog},V_{Cloud},C_{Cloud}\) are respectively the amount of data in Fog, the computation power of Fog, the amount of data in Cloud, and the computation power of Cloud. As discussed in Section II, the computation power of Fog is limited. Therefore, after a certain amount of data, the BS will send the data directly to the Cloud. In this simulation, we have considered the computation power of the Fog as 1GB/cycle, which means it can compute the 1GB of data per cycle. Finally, using the above equations, we define Total Latency as \(T_{total}\) equal to the summation of the Latency for every UEs from 1 to N considering the Processing Time and the double of Transmission Time caused by sending data and receiving the processed information.

\begin{equation*}
T_{total}=\sum_{i=1}^{N}T_{T}^{i}+T_{P}^{i}
\end{equation*}
\begin{equation} \label{eq1}
T_{T}^{i}=2(\frac{\alpha V_{Data}^{i}}{S_{UE}^{i}}+\frac{V_{Fog}^{i}}{S_{BS}}+\ln{H}*\frac{V_{Cloud}^{i}}{S_{BS}})
\end{equation}
\begin{equation*}
T_{P}^{i}=\frac{V_{Fog}^{i}}{C_{Fog}}+\frac{V_{Cloud}^{i}}{C_{Cloud}}, V_{Data}^{i}=V_{Fog}^{i}+V_{Cloud}^{i}
\end{equation*}

In the second model, we simulate a simple 5G MEC consisting of four types of agents: UE, 5G gNB, MEC, and Cloud as shown in Fig \ref{fig:6} (b). The UE continuously sends data to 5G gNB that is responsible to insert the generated data in MEC or send it to Cloud for more detailed computation. Same as the first simulation, in this simulation we measure the latency, which is the time between sending the raw data and receiving the information. As shown in (2), the Transmission Time of this simulation is equal to the sum of the time that data is sent and received from UE to 5G gNB, MEC, and Cloud. The ratio speed of 5G or \(S_{5GgNB}\) has been estimated 800 megabytes per second and the motion effect of the UE or \(\alpha\) is defined as a factor between 0.8 and 1 that multiplies in the speed rate. The same as the previous model, the distance between UE-to-5G\_gNB and 5G\_gNB-to-MEC is set to a single hop, which has a linear effect on the latency considering the improvement of 5G to compare with 4G LTE. Also, the distance between 5G gNB and Cloud is considered equal to 10 hops. This distance slightly affects the latency due to the power loss and the capacity of the 5G communication channel. However, it is still possible to reduce the latency of 5G MEC as compared with the proposed hybrid model. On the other hand, Processing Time is equal to the time that data is processed by MEC and Cloud.  Moreover, as shown in Section II, the computation power of MEC is superior to Fog and it has been considered equal to 2GB/cycle in this simulation.
\begin{figure*}[h]
    \centering
    \includegraphics[scale=0.4]{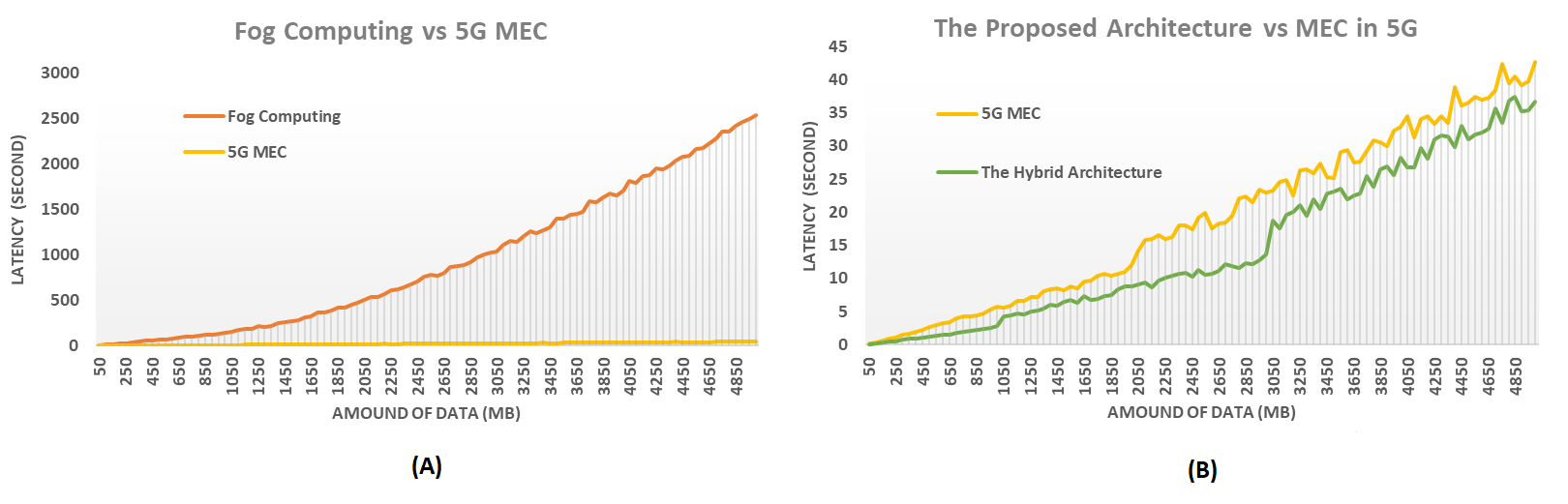}
    \caption{\textbf{Comparison Between Latencies of the Simulated Architectures}}
    \label{fig:7}
 \end{figure*}
\begin{equation} \label{eq2}
T_{T}^{i}=2(\frac{\alpha V_{Data}^{i}}{S_{UE}^{i}}+\frac{V_{MEC}^{i}}{S_{5GgNB}}+\ln{H}*\frac{V_{Cloud}^{i}}{S_{5GgNB}})
\end{equation}
\begin{equation*}
T_{P}^{i}=\frac{V_{MEC}^{i}}{C_{MEC}}+\frac{V_{Cloud}^{i}}{C_{Cloud}}, V_{Data}^{i}=V_{MEC}^{i}+V_{Cloud}^{i} 
\end{equation*}
In the last simulation, we have simulated the proposed hybrid architecture that consists of 6 types of agents: UE, 5G CU-CP, 5G CU-UP, C-MEC, U-MEC, and Cloud as shown in Fig \ref{fig:6}(c). The strategy is that the UE continuously sends data to 5G CU-UP that is responsible to insert the data in U-MEC or to send it to the 5G CU-CP for processing the rest of the data in C-MEC or either send it to Cloud for more detailed computation. Same as the first and the second simulation, we measure the time between sending the raw data and receiving the information as latency in this simulation. The communication system between UEs and other components is based on 5G and its speed rate is defined 800 megabytes per second. The motion of the UE is defined as a factor between 0.9 and 1 that multiplies in the speed rate. The same as the previous model, the distance between UE-to-5G-CU-UP and 5G-CU-to-MEC is set to a single hop, which has a linear effect on the latency considering the improvement of 5G to compare with 4G LTE.  Same as 5G MEC, the distance between the 5G CU-CP and the Cloud is considered equal to 10 hops, which has a nonlinear effect on the latency considering the power loss and the capacity of the communication channel as well. The computation power of U-MEC and C-MEC is considered 2GB/Cycle and 1GB/Cycle respectively.
\begin{equation*} 
\frac{T_{T}^{i}}{2}=\frac{\alpha V_{Data}^{i}}{S_{UE}^{i}}+\frac{V_{U-MEC}^{i}}{S_{5G-CU-U}}+\frac{V_{C-MEC}^{i}}{S_{5G-CU-C}}+\ln{H}*\frac{V_{Cloud}^{i}}{S_{5G-CU}}
\end{equation*}
\begin{equation}\label{eq3}
T_{P}^{i}=\frac{V_{U-MEC}^{i}}{C_{U-MEC}}+\frac{V_{C-MEC}^{i}}{C_{C-MEC}}+\frac{V_{Cloud}^{i}}{C_{Cloud}}
\end{equation}
\begin{equation*}
V_{Data}^{i}=V_{U-MEC}^{i}+V_{C-MEC}^{i}+V_{Cloud}^{i} 
\end{equation*}
As shown in Fig \ref{fig:7}, we have run the multi-agent systems and fed them by 50 MB data to 2.5 GB of input with a ratio of 50MB. The latency of the Fog Computing paradigm is not comparable to the latency of 5G MEC as shown in Fig \ref{fig:7}(A). The main reason for this discrepancy is that the speed rate of 5G is 200 times more than 4G LTE-Advanced. We also have simulated the proposed hybrid model and compare its latency with the latency of 5G MEC. As shown in Fig \ref{fig:7}(B), the proposed model reduces latency up to 11\% in comparison with the latency of 5G MEC. The main reason for the difference is that there are more available resources close to the UEs. Moreover, the network has been customized by SDN controllers to offload the resources distributed in certain areas.

In this section, we have performed an agent-based simulation and compared the result with the Fog Computing paradigm and 5G MEC paradigm. We demonstrated that the proposed architecture has less latency comparing 5G MEC. Finally, we have discussed the result to prove and show the contribution of our paper. Besides the latency, there are other performance measurements such as operation cost, QoS, and energy efficiency that have to be compared with 5G MEC in the implementation phase. In the next section, we will discuss an open-source project in order to implement the model practically with available tools and also about the open issues that need to be considered and investigated.

\section{Discussion}
 So far, we have proposed an architecture for 5G Edge Computing in order to design a solution that satisfies the requirements of the delay-sensitive applications. In this section, we are going to compare the proposed architecture with similar hybrid design architectures for Edge Computing. Afterward, we will discuss about the open issues that need to be considered in the proposed architecture. These issues can provide direction to researchers for further research in this area.

\begin{table*}[h]
\begin{center}
\caption{Comparison Between Hybrid Architectures for Edge Computing}
\begin{tabular}{ | P{4cm} | P{1.5cm} | P{1.5cm} | P{2cm} | P{2cm} | } 
\hline
Hybrid Architectures &Scalability&Reliability&ULL Support&Success Rate\\
\hline
Li et al. \cite{b58}&Low&Medium&Low&Medium\\ 
\hline
A. Reiter et al. \cite{b59}&Medium&Medium&Medium&Medium\\ 
\hline
L. Pu et al. \cite{b60}&Low&Medium&High&Medium\\ 
\hline
The Proposed Architecture &	High&High&High&High \\ 
\hline
\end{tabular}
\end{center}
 \end{table*}
\subsection{Comparison} 
There are various traditional designed hybrid architectures for Edge Computing. Most of these architectures have not considered the requirements of ultra-low latency applications and the next-generation technologies that can satisfy them. In Table V, we have compared these architectures with each other based on the criteria we introduced as the benefits of Edge Computing in Section II. For instance, although Li et al. \cite{b58} proposed a four-layer hybrid computing framework to fulfill the real-time requirements of smart manufacturing with edge computing support, they didn't consider the structure of standardized deployment of edge computing, such as MEC or Cloudlets. Also, their research is based on current technologies. For instance, they didn't investigate the next-generation networks such as 5G or beyond the 5G. In \cite{b59}, A. Reiter et al. proposed a novel architecture for Hybrid Mobile Edge Computing by dynamically using the edge-computing-provided computational power in order to overcome battery limitations and performance constraints.  Although their research has considered the improvement of performance and energy consumption of Edge Computing, their proposed architecture is not scalable for ultra-low latency applications, which require fast communication and ultra-speed data processing with next-generation technologies such as Massive MIMO or beyond the 5G networks. In \cite{b60}, L. Pu et al. proposed a novel hybrid edge computing framework for future large-scale vehicular crowdsensing applications to augment the resources of connected vehicles.

\subsection{Open Research Issues}
In this section, we are going to discuss about the open issues in the deployment of 5G Edge Computing\cite{b61}\cite{b62}:
\begin{itemize}
  \item \textbf{Quality of Experience (QoE):} QoE measures the level of the customer’s satisfaction with the provided service. QoE is different from QoS and embodies the notion that the hardware and software characteristics could be guaranteed and enhanced
  \item \textbf{Interoperability and Standardization:} Standardization requires related organizations to provide a set of universally acceptable rules for 5G Edge Computing. The diversion of customization by different vendors has made it difficult to agree upon a standard.  Billions of devices are going to use various interfaces for communication between UEs and edge clouds. Therefore, organizations such as the European Telecommunications Standards Institute (ETSI) have to be formed in order to address interoperability issues.
  \item \textbf{Security and Privacy:} In order to address security issues such as vulnerability at the edge of the network, we need a scalable solution that manages trust and privacy concerning complexity and cost.
\end{itemize}

\section{Conclusion}
Edge Computing in 5G and Beyond is a promising solution for ultra-low latency applications that are sensitive to delay and require fast data processing. In this paper, we profoundly studied Edge Computing and investigated the role of 5G in Edge Computing. Then, we proposed a hybrid architecture, consisting of MEC and Edge Cloud, based on the requirements of delay-sensitive use cases of 5G Edge Computing. In order to evaluate the proposed architecture, we performed an agent-based simulation on the three models, Fog Computing, 5G MEC, and the proposed architecture, and demonstrated that the proposed architecture decreases the latency up to 11\% compared to 5G MEC. Finally, as the conclusion, we compared the proposed architecture with similar hybrid designed architectures for Edge Computing, and discussed the open issues while looking at the bigger picture.

\end{document}